\documentclass[letterpaper, 12 pt, conference]{ieeeconf}  % Comment this line out
\IEEEoverridecommandlockouts                              % This command is only
\usepackage{graphicx} % for pdf, bitmapped graphics files
\usepackage{bm}

\usepackage{epsfig} % for postscript graphics files
\usepackage{mathptmx} % assumes new font selection scheme installed
\usepackage{times} % assumes new font selection scheme installed
\usepackage{amsmath} % assumes amsmath package installed
\usepackage{amssymb}  % assumes amsmath package installed

\title{\LARGE \bf 
An unlikely route to low lattice thermal conductivity: small atoms in a simple layered structure}
\author{Wanyue Peng \textit{$^{a}$}, Guido Petretto \textit{$^{b}$}, Gian-Marco Rignanese \textit{$^{b}$}, \\ Geoffroy Hautier \textit{$^{b}$}, Alexandra Zevalkink \textit{$^{a}$}% <-this % stops a space
\thanks{}% <-this % stops a space
\thanks{\textit{$^{a}$~Michigan State University, East Lancing, MI, USA.}}%
\thanks{\textit{$^{b}$~Universit\'e catholique de Louvain, Louvain-la-Neuve, Belgium}}%
}

\begin{document}

\maketitle
\thispagestyle{empty}
\pagestyle{empty}

%%%%%%%%%%%%%%%%%%%%%%%%%%%%%%%%%%%%%%%%%%%%%%%%%%%%%%%%%%%%%%%%%%%%%%%%%%%%%%%%
\begin{abstract}

In the design of materials with low lattice thermal conductivity, compounds with high density, low speed of sound, and  complexity at either the atomic, nano- or microstructural level are preferred.  
%ET: Maybe mention that scattering sources are also important
The layered compound Mg$_3$Sb$_2$ defies these prevailing paradigms, exhibiting lattice thermal conductivity comparable to PbTe and Bi$_2$Te$_3$, despite its low density and simple structure.  The excellent thermoelectric performance ($zT$ $\sim$ 1.5) in $n$-type Mg$_3$Sb$_2$ has thus far been attributed to its multi-valley conduction band, while its anomalous thermal properties have been largely overlooked.  To explain the origin of the low lattice thermal conductivity of Mg$_3$Sb$_2$, we have used both experimental methods and ab initio phonon calculations to investigate trends in the elasticity, thermal expansion and anharmonicity of $A$Mg$_2Pn_2$ Zintl compounds with $A$ = Mg, Ca, Yb, and $Pn$ = Sb and Bi.  Phonon calculations within the quasi-harmonic approximation reveal large mode Gr\"uneisen parameters in Mg$_3$Sb$_2$ compared with isostructural compounds, in particular in transverse acoustic modes involving shearing of adjacent anionic layers. Measurements of the elastic moduli and sound velocity as a function of temperature using resonant ultrasound spectroscopy provide a window into the softening of the acoustic branches at high temperature, confirming their exceptionally high anharmonicity. We attribute the anomalous thermal behavior of Mg$_3$Sb$_2$ to the diminutive size of Mg, which may be too small for the octahedrally-coordinated site, leading to weak, unstable interlayer Mg-Sb bonding.  This suggests more broadly that soft shear modes resulting from undersized cations provide a potential route to achieving low lattice thermal conductivity low-density, earth-abundant materials.
\end{abstract}

%%%%%%%%%%%%%%%%%%%%%%%%%%%%%%%%%%%%%%%%%%%%%%%%%%%%%%%%%%%%%%%%%%%%%%%%%%%%%%%%
\section{INTRODUCTION}
The ability to predict and design thermal transport in bulk materials is a fundamental requirement in a wide range of applications.  In areas such as the development of thermal barrier coatings and thermoelectric materials, engineering materials with extremely low lattice thermal conductivity, $\kappa_L$, is vital.  In materials with inherently high thermal conductivity (e.g., Si), low $\kappa_L$ can be achieved using clever microstructural design, nanostructuring, defects, or alloying to scatter phonons \cite{biswas2012high,bux2010nanostructured,klemens1955scattering}. Alternatively, inherently low $\kappa_L$ can be achieved in compounds with either low phonon velocities or high rates of phonon-phonon Umklapp scattering. Thus, in the search for materials with low $\kappa_L$, compounds with high density (PbTe, Bi$_2$Te$_3$), soft bonds, and complex atomic structures (e.g., Yb$_{14}$MnSb$_{11}$ with 104 atoms per unit cell \cite{toberer2008traversing}) are favored, since these features lead to low phonon velocities \cite{toberer2011phonon}. Design parameters for achieving high rates of phonon-phonon scattering (i.e., materials with large Gr\"uneisen parameters) are more elusive, although a number of studies have recently provided guidance here as well.  For example, highly anharmonic phonon modes arise from soft or unstable bonds (e.g., associated with Cu lone pairs, rattlers in cage compounds, resonance bonding, etc. \cite{lee2014resonant,madsen2005anharmonic,lai2015bonding}) 
and they generally emerge in the vicinity of lattice instabilities \cite{bansal2016phonon,delaire2011giant}.

%ET: cite Toberer EES search for low lattice thermal conductivity 

\begin{figure}[!ht]
    \centering
    \includegraphics[width=0.45\textwidth]{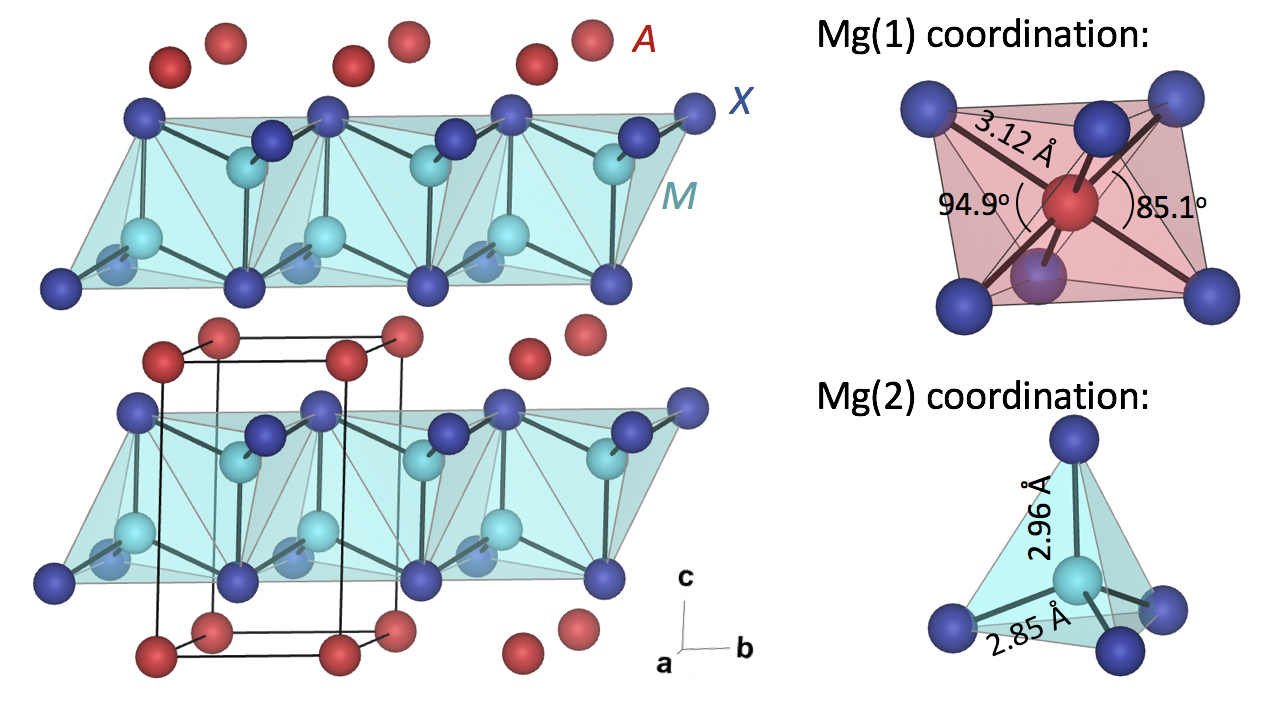}
    \caption{Mg$_3$Sb$_2$ crystallizes in the CaAl$_2$Si$_2$ structure (space group $P\bar{3}m1$), characterized by anionic $M_2X_2$ slabs separated by $A$ cations. In the binary compounds Mg$_3$Sb$_2$ and Mg$_3$Bi$_2$, the Mg(1) occupies the octahedrally-coordinated $A$ site and Mg(2) the tetrahedrally-coordinated $M$ site \cite{martinez1974crystal}. }
    %Let's cite the original Zintl Mg3Sb2 paper here from the 1920's.(1920s? I think this 1974 one is better.)
   \label{fig:structure}
\end{figure}

\begin{figure*}[!ht]
\centering
\includegraphics[width=0.9\textwidth]{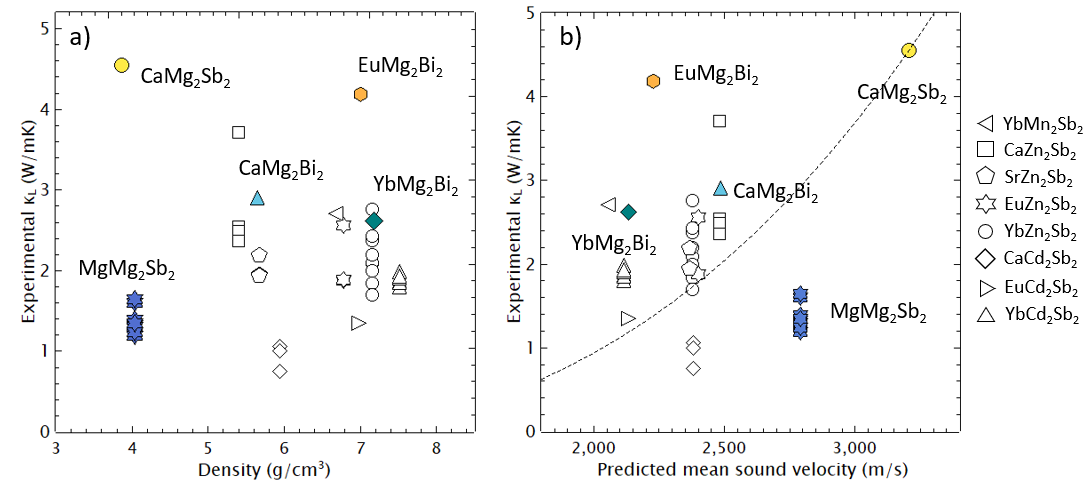}
\caption{The experimental lattice thermal conductivity, $\kappa_L$, of Mg$_3$Sb$_2$ is significantly lower than isostructural  $AM_2X_2$ compounds with similar a) density and b) predicted speed of sound ($\kappa_L$ data can be found in Ref. \cite{122review}). The mean speed of sound, $v_s$, was estimated using the calculated elastic moduli from MaterialsProject.org and experimental densities \cite{ceder2010materials,de2015charting}.  The dashed line shows a $\kappa_L \propto v_s^3$ dependence as a guide to the eye.  Compounds with $M$ = Mg are shown in color, as these are the primary focus of the current study.}
\label{fig:kappa}
\end{figure*}
% ET:  Make this caption into an "Action caption".  It's really boring right now.  Reference side b) use bigger text, explain why M=Mg ones are colored

The binary compound, Mg$_3$Sb$_2$, which crystallizes in the layered CaAl$_2$Si$_2$ structure type, appears to defy standard paradigms used to identify materials with low lattice thermal conductivity.  Mg$_3$Sb$_2$ has a low density, relatively high speed of sound, and a simple atomic structure with only five atoms per primitive cell, yet despite having roughly half the density of PbTe and Bi$_2$Te$_3$, it has comparable $\kappa_L$ at room temperature (1-1.5 W/mK \cite{bhardwaj2013mg,song2017simultaneous,shuai2015thermoelectric}).  As illustrated by Fig. \ref{fig:kappa}a, the low $\kappa_L$ of Mg$_3$Sb$_2$ is anomalous compared with other $AM_2X_2$ compounds with the CaAl$_2$Si$_2$ structure type. Although Mg$_3$Sb$_2$ is one of the lightest compounds in the series, it exhibits one of the lowest values of $\kappa_L$.  In particular, we note that the reported $\kappa_L$ of Mg$_3$Sb$_2$ is much lower than that of CaMg$_2$Sb$_2$ [cite Max], which differs only by the presence of Ca instead of Mg on the octahedral site.
%ET: This paragraph is weak and kind-of weaves in and out.  Maybe make the point that 122's are not known for low kappa_L?

$AM_2X_2$ compounds with the CaAl$_2$Si$_2$ structure have attracted a great deal of interest for thermoelectric applications \cite{shuai2017recent,122review} owing to the chemical diversity and flexibility in tuning transport properties.  Most commonly, $A$ is divalent cation such as Mg, Ca, Yb, Eu, $M$ is a divalent metal such as Mg, Mn, Zn, or Cd, and $X$ is a pnictogen species. Shown in Figure \ref{fig:structure}, the structure of CaAl$_2$Si$_2$ is composed of 2-dimensional anionic $M_2X_2$ slabs in which $M$ is tetrahedrally coordinated by $X$ \cite{burdett1990fragment}. The cations occupying the layers between the anion slabs are octahedrally coordinated by $X$.  In the binary members of this structure (e.g., Mg$_3$Sb$_2$ and Mg$_3$Bi$_2$), the Mg(1) occupies the octahedral $M$ site and Mg(2) the tetrahedral $A$ site, and exhibits significantly different bond length at each site.

Recently, excellent thermoelectric performance has been demonstrated in $n$-type Mg$_3$Sb$_2$-based samples, with $zT$ up to 1.6 reported by several independent groups \cite{tamaki2016isotropic, zhang2017high, Iverson2017_ncomms_n_Mg2Sb3, imasato2017band, shuai2017tuning}. This surpasses all previous results for isostructural compounds, all of which have been $p$-type.   To date, experimental and theoretical investigations of Mg$_3$Sb$_2$ have focused on the electronic properties; e.g., the defect origin of $n$-type doping \cite{ohno2017phase,tamaki2016isotropic}, the multi-valley character of the conduction band \cite{zhang2017high,Zhang2015,shuai2016higher,imasato2017band}, and routes to increased carrier mobility \cite{Kuo2018grain,shuai2017tuning}.  In contrast, the anomalously low $\kappa_L$ of both $n$- and $p$-type Mg$_3$Sb$_2$ has not been investigated, though it plays an equally important role in leading to the high $zT$.   The aim of the present study is to shed light on the origins of low lattice thermal conductivity in Mg$_3$Sb$_2$.  We employ experimental methods and ab initio phonon calculations to investigate trends in the thermal properties of binary and ternary $A$Mg$_2Pn_2$ compounds with $A$ = Mg, Ca, Yb, and $Pn$ = As, Sb and Bi, revealing previously unrecognized soft shearing modes and highly anharmonic acoustic phonons in Mg$_3$Sb$_2$. This work shows that the soft shear modes resulting from undersized cations provide a potential route to achieving low lattice thermal conductivity in simple, low-density structures.

%%%%%%%%%%%%%%%%%%%%%%%%%%%%%%%%%%%%%%%%%%%%%%%%%%%%%%%%%%%%%%%%%%%%%%%%%%%%%%%%%%%%%%%%%%%%%%%%%%%%%%%%%%%%%%%%%%%%%%%%%%%%%%%%%%%%%%%%%%%%%%%%%%%%%%%%%%%%%%%%%%%%%%%%%%%%%%%%%%%%%%%%%%%%%%%%%%%%%%

\section{Methods}

\subsection{Synthesis}
$A$Mg$_2$$Pn$$_2$ compounds with $A$=Mg, Ca, Yb and $Pn$=Sb, Bi were synthesized by direct ball-milling of the elements followed by spark plasma sintering. The corresponding stoichiometric elements (99.8\% Mg shot, 99.5\% Ca shot, 99.9\% Yb chunk, 99.99\% Sb from Alfa Aesar and 99.99\% Rotometal Bi) were cut into small pieces in an argon filled glovebox, loaded into stainless steel vials with two 10 mm diameter stainless balls, and milled for one hour using a SPEX mill. The powder was then loaded into graphite dies with 10 mm bores and sintered using the profile shown in Table \ref{table_SPS} under a pressure of 31 MPa using a Dr. Sinter SPS-211LX. The pressure was removed immediately when cooling started. The densities of all the samples were obtained by measurement of mass and geometry. All samples were at least 97\% of the theoretical density. Phase purity was confirmed using a Rigaku X-ray Diffraction system, showing that samples contained less than 3\% of secondary phases.

\begin{table}[!ht]
\centering
\caption{Maximum temperature and hold time used during spark plasma sintering of $A$Mg$_2$$Pn$$_2$ ($A$=Mg, Ca, Yb and $Pn$=Sb, Bi) samples.}
\label{table_SPS}
\begin{tabular}{lccc}
\hline
\multicolumn{1}{c}{} & Mg$_3$Sb$_2$ & CaMg$_2$Sb$_2$ & YbMg$_2$Sb$_2$  \\
Temp. ($^{\circ}C$)  & 850      & 650        & 650           \\
Time (min)   & 15      & 10        & 10        \\   \multicolumn{1}{c}{}  & Mg$_3$Bi$_2$ & CaMg$_2$Bi$_2$ & YbMg$_2$Bi$_2$ \\
Temp. ($^{\circ}C$)         & 600      & 700        & 700        \\
Time (min)      & 10     & 10      & 15  \\   
\hline
\end{tabular}
\end{table}

\subsection{Characterization}
The temperature-dependent elastic moduli of all compositions were measured by resonant ultrasound spectroscopy (RUS) \cite{li2010high,astm1876standard} using a custom modification of a Mangaflux-RUS Quasar 4000 system in a furnace with a flowing Ar atmosphere. Cylindrical samples were mounted on a tripod transducer setup. One transducer induced the mechanical vibrations and the remaining two detected the specimen resonances. The elastic moduli were measured in 20 K intervals from 303 K up to 673 K. The sinusoidal driving frequency was swept from 0 to 500 kHz. The RUS measurement at each temperature was typically completed within one minute. Data was analyzed using commercial Quasar2000 CylModel software to match predicted and observed resonant frequencies.

\begin{figure*}[!ht]
    \centering
    \includegraphics[width=0.8\textwidth]{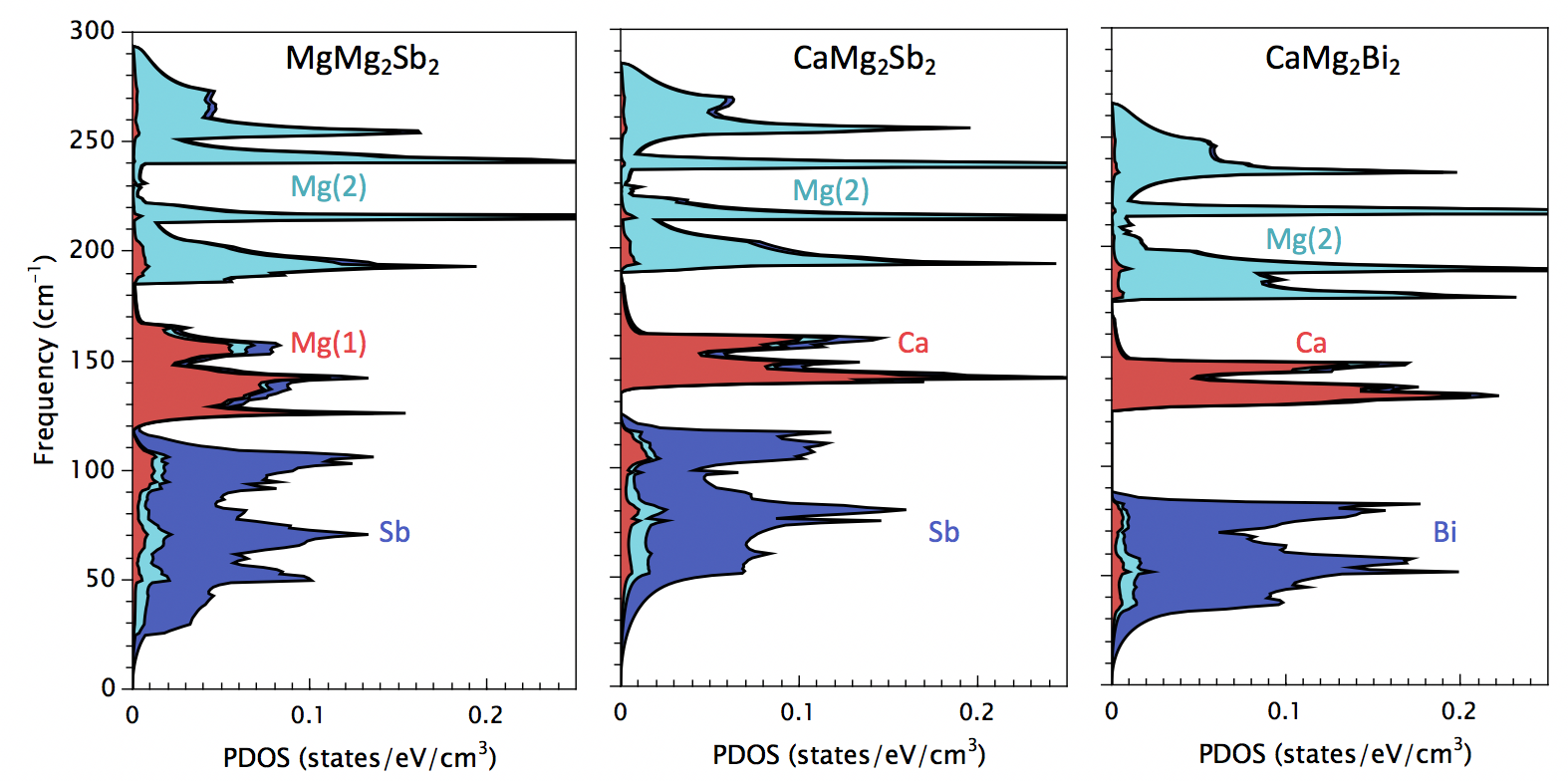}
    \caption{The atom-projected phonon density of states of Mg$_3$Sb$_2$, CaMg$_2$Sb$_2$ and CaMg$_2$Bi$_2$ shows that the low-, mid-, and high- frequency regimes are dominated, respectively, by displacements of the anions (Sb, Bi), cations (Mg(1), Ca), and metal site (Mg(2)).  The species on the cation site (Mg or Ca) does not strongly influence the maximum frequency or the average phonon velocities.}
   \label{fig:DOS}
\end{figure*}
%ET: mention this is a stacked DOS.

%\paragraph{High-temperature X-ray diffraction (HT-XRD)}
Thermal expansion of all compositions was measured from 303 K to 573 K using a Rigaku Smartlab XRD equipped with a high-temperature stage. The samples were ground into fine powders that were then placed on a graphite foil on top of a a platinum tray. The measurements were performed under vacuum to prevent oxidation. The thermocouple goes into the inner part of the platinum tray to increase the accuracy of the temperature measurement. The heating rate is 10 K/min with a 1 min hold and sample height alignments were performed before each measurement to account for the combined thermal expansion of the holder and sample.

\subsection{Calculation details}
The ABINIT software package\cite{abinit2005,abinit2009,abinit2016} was used to perform density functional theory (DFT) and density functional perturbation theory (DFPT) simulations to obtain phonon properties and elastic constants \cite{BaroniDFPT2001, GonzeDFPT1997a, GonzeDFPT1997b, Hamann2005}. The exchange-correlation functional was described using the PBEsol\cite{PBEsol} approximation, that has proven to provide accurate phonon frequencies compared to experimental data\cite{He2014}. Norm-conserving pseudopotentials\cite{Hamannn2013} extracted from the Pseudo-dojo pseudopotentials table version 0.3\cite{pseudodojo} were used for all the elements and the Brillouin zone was sampled with $8\times8\times5$ Monkhorst-Pack grids \cite{Monkhorst1976, petretto2018a}. 
% add petretto2018b citation later.  It's not published yet.
Due to standard DFT's known limitations in describing the band gap, we limit our analysis to Mg$_3$Sb$_2$, CaMg$_2$Sb$_2$ and CaMg$_2$Bi$_2$, for which the electronic structures are correctly described in our approximations.  Temperature-dependent thermal expansion data was extracted from the phonon dispersion by relaxing the position of the atoms and shape of the unit cell at different fixed volumes in the framework of the quasiharmonic approximation. Once extracted the thermal expansion coefficient, the Young's modulus, $Y$, was calculated as a function of temperature by using the corresponding unit cell volume, from which 300 K was chosen as the initial point $Y_o$. Gr\"{u}neisen parameters were obtained as the derivative of the phonon frequencies with respect to the volume.

%%%%%%%%%%%%%%%%%%%%%%%%%%%%%%%%%%%%%%%%%%%%%%%%%%%%%%%%%%%%%%%%%%%%%%%%%%%%%%%%%%%%%%%%%%%%%%%%%%%%%%%%%%%%%%%%%%%%%%%%%%%%%%%%%%%%%%%%%%%%%%%%%%%%%%%%%%%%%%%%%%%%%%%%%%%%%%%%%%%%%%%%%%%%%%%%%%%%%%

\section{Results and discussion}

\subsection{Anomalously low lattice thermal conductivity in Mg$_3$Sb$_2$}

Inherently low lattice thermal conductivity, $\kappa_L$, stems from either slow phonon velocities, $v$, or short phonon relaxation times, $\tau$ \cite{slack1979thermal,toberer2011phonon}.  Of these two factors, the phonon velocities are the more accessible quantity; they can either be estimated roughly using the speed of sound, or they can be obtained from the calculated or measured phonon dispersion. Thus, in investigating the origin of low $\kappa_L$ in any material, we should always begin by asking whether or not low phonon velocity is responsible.   In Figure \ref{fig:kappa}b, we have plotted the experimental $\kappa_L$ of $AM_2X_2$ compounds as a function of the predicted mean speed of sound, $\nu_s$.  We have only included data from unalloyed samples, most of which exhibit the 1/T temperature dependence expected for Umklapp scattering dominated transport.  Due to the lack of experimental speed of sound data in most of the compounds, we used the calculated bulk and shear elastic moduli from Materials Project to estimate $\nu_s$ \cite{de2015charting} (details can be found in S.I. section 1).  
A comparison of the experimental and computed speed of sound in the compounds in this study show less than 7\% difference, showing the reliability of this approach. 

When thermal transport is limited by Umklapp phonon-phonon scattering, $\kappa_L$ can be approximated as proportional to $\nu_v^3/\gamma^2$, where $\gamma$ is the Gr\"uneisen parameter \cite{slack1979thermal,toberer2011phonon,zeier2017new}.  The dashed curve in Figure \ref{fig:kappa}b has a $\nu_s^3$ dependence, provided as a ``guide to the eye" to show that the low $\kappa_L$ of Mg$_3$Sb$_2$ cannot be explained by taking into account the sound velocity.  This implies either a) that additional scattering sources are present and unique to Mg$_3$Sb$_2$ samples, b) that the Gr\"uneisen parameter of Mg$_3$Sb$_2$ is abnormally large, leading to increased Umklapp scattering or c) that the sound velocity fails to capture broader trends in the phonon group velocities in this structure type.  We note that there is currently no evidence that the microstructure or defect concentrations in Mg$_3$Sb$_2$ samples differ strongly from other $AM_2X_2$ samples. Similar values for $\kappa_L$ reported from multiple research groups and processing approaches lead us to believe that low $\kappa_L$ is an inherent characteristic of Mg$_3$Sb$_2$ \cite{bhardwaj2013mg,Iverson2017_ncomms_n_Mg2Sb3,song2017simultaneous,shuai2015thermoelectric,shuai2017tuning,ohno2017phase}.  Thus, in the current investigation, we focus on the latter two inherent explanations for low $\kappa_L$.

%%%%%%%%%%%%%%%%%%%%%%%%%%%%%%%%%%%%%%%%%%%%%%%%%%%%%%%%%%%%%%%%%%%%%%%%%%%%%%%%%%%%%%%%%%%%%%%%%%%%%%%%%%%%%%%%%%%%%%%%%%%%%%%%%%%%%%%%%%%%%%%%%%%%%%%%%%%%%%%%%%%%%%%%%%%%%%%%%%%%%%%%%%%%%%%%%%%%%%

\begin{figure}[!ht]
    \centering
    \includegraphics[width=0.45\textwidth]{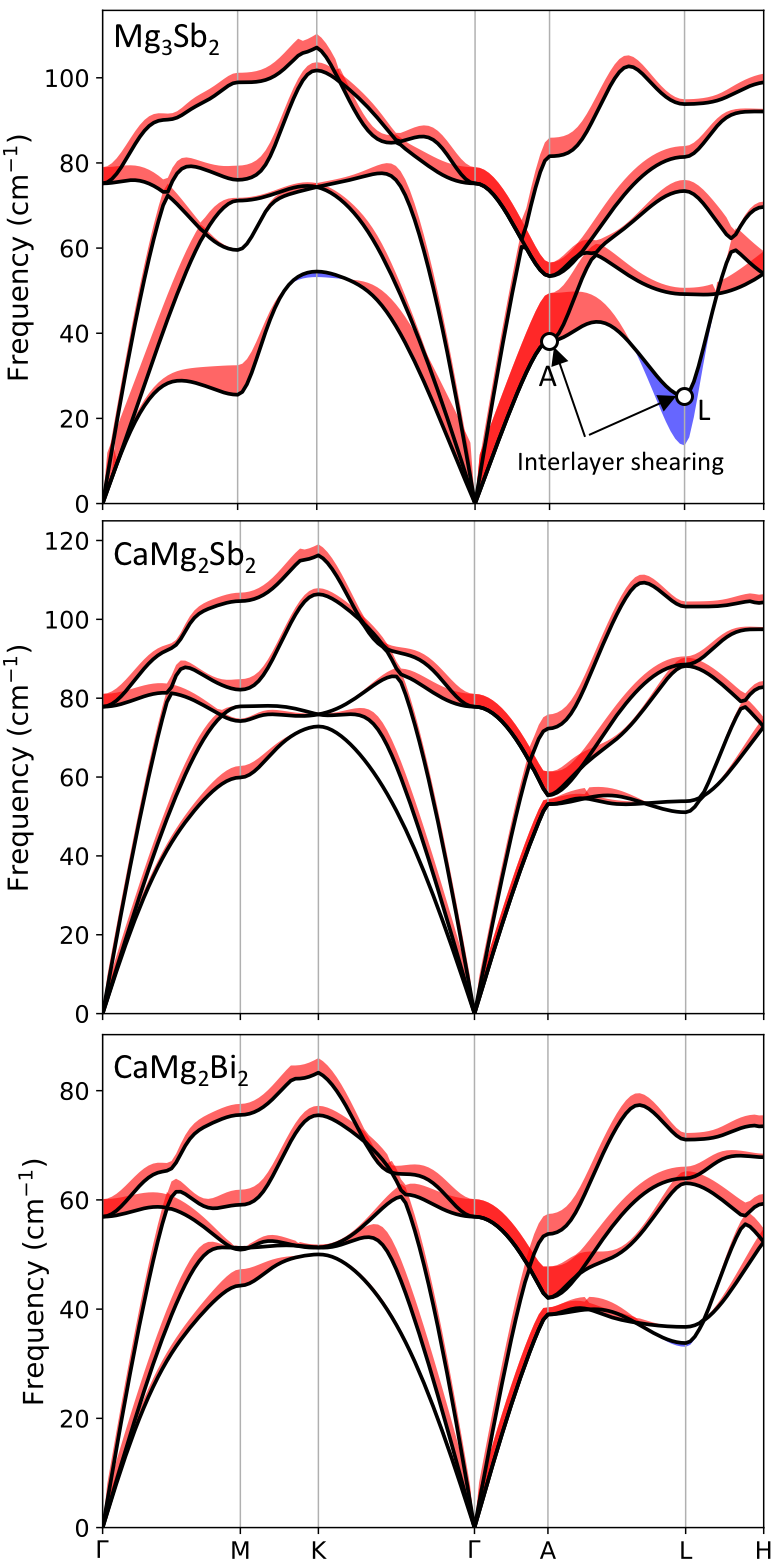}
    \caption{The phonon dispersions of Mg$_3$Sb$_2$, CaMg$_2$Sb$_2$ and CaMg$_2$Bi$_2$ in the low frequency regime. The mode Gr\"uneisen parameters, $\gamma_i$, are shown through the thickness of the bands, with red and blue representing positive and negative values of $\gamma_i$, respectively.}
   \label{fig:dispersion}
\end{figure}
%ET: consider adding the total phonon DOS next to the dispersion, mention the spike near Sb

\begin{figure}[!ht]
    \centering
    \includegraphics[width=0.48\textwidth]{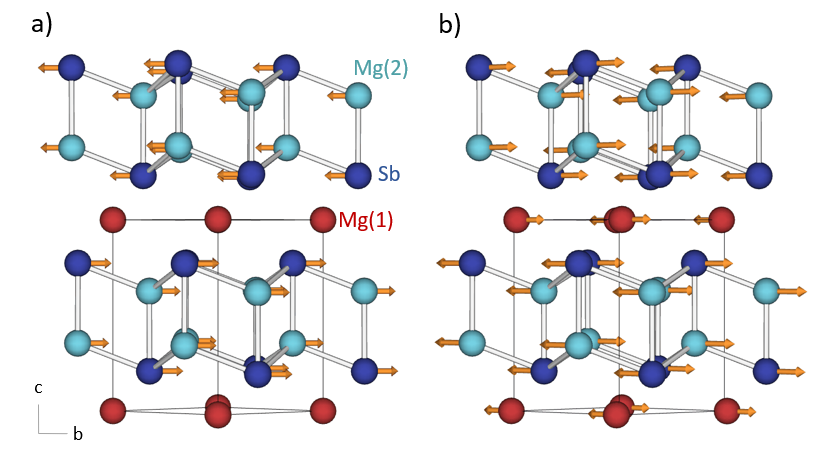}
    \caption{Structure of Mg$_3$Sb$_2$ showing the atomic displacement corresponding to the transverse acoustic phonon modes a) at the $A$-point (large positive $\gamma_i$) and b) at the $L$-point (large negative $\gamma_i$).  The corresponding modes are marked on the phonon dispersion in Figure \ref{fig:dispersion}.
}
\label{fig:displacement}
\end{figure}

\subsection{Ab initio phonon calculations}

\begin{figure*}[!ht]
    \centering
    \includegraphics[width=0.8\textwidth]{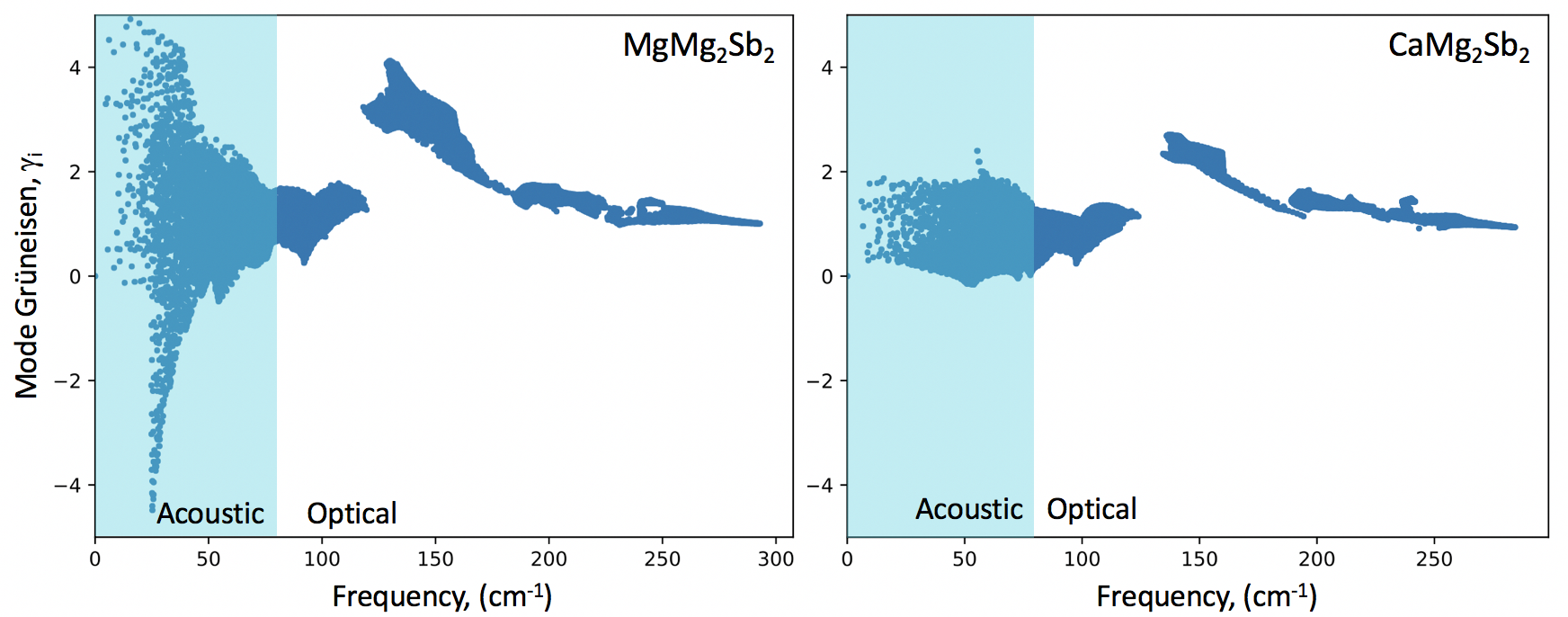}
    \caption{The mode Gr\"uneisen parameters as a function of frequency highlight the impact that the cation site (Mg versus Ca) has on the low frequency, acoustic phonons.  In contrast, optical phonons are not strongly impacted. Data for CaMg$_2$Bi$_2$, which behaves similarly to CaMg$_2$Sb$_2$, is shown in the supplemental.}
   \label{fig:gruneisen}
\end{figure*}

% consider adding a Brillouin zone figure to show different directions

To date, the calculated phonon density of states (DOS) and dispersion relations have only been reported for a small handful of compounds with the CaAl$_2$Si$_2$ structure type \cite{singh2013electronic, tani2010lattice}.  The atom-projected DOS shown for Mg$_3$Sb$_2$, CaMg$_2$Sb$_2$, and CaMg$_2$Bi$_2$ in Figure \ref{fig:DOS}, are consistent with previous reports, which show sharply segmented frequency regimes. The anion displacement ($Pn$ = Sb or Bi) dominates at low frequencies, cation displacement is responsible for the mid-frequency modes, and the $M$=Mg site dominates the highest frequency range.  In the case of Mg$_3$Sb$_2$, the partitioning of the two Mg sites reflects very significant differences in local bonding environment.  From the lower phonon frequencies of the octahedrally-coordinated Mg(1), we can infer that the Mg(1)-Sb bonds are much softer than the Mg(2)-Sb bonds.  In fact, Mg(1) occupies a similar frequency range to that Ca, showing that its bonding environment is more like that of Ca than that of Mg(2).
%ET: Emphasize the DOS better.  This shows that Mg(1) behaves like Ca, while Mg(2) behaves the same.  Makes it more Zintl like

Acoustic phonons tend to have an out-sized influence on thermal transport due to the strong frequency dependence of Umklapp ($\tau \propto 1/\omega^2$) and point defect scattering ($\tau \propto 1/\omega^4$) \cite{tritt2005thermal,toberer2011phonon}, which leads to long mean free paths for low frequency phonons.
The acoustic branches also tend to have higher group velocities than the optical branches, which further amplifies their relative contribution to $\kappa_L$.  Any mechanism that preferentially slows or scatters acoustic phonons, therefore, can have an enormous impact on $\kappa_L$.  Given that the acoustic phonons are dominated by displacement of the anions (Figure \ref{fig:DOS}), it is not immediately clear why the cation species (Mg or Ca) would greatly impact the phonon transport.  Further, the average phonon group velocities at high frequencies and the Debye frequency of Mg$_3$Sb$_2$ do not appear significantly different from CaMg$_2$Sb$_2$.  We can therefore rule out differences in the phonon group velocity as the $primary$ origin of the low lattice thermal conductivity in Mg$_3$Sb$_2$.

The major differences between CaMg$_2$Sb$_2$ and Mg$_3$Sb$_2$ first become clear when we consider the phonon dispersion relations and their dependence on unit cell volume.  The phonon dispersions are shown in Figure \ref{fig:dispersion} in the low frequency range only, as these show the greatest change with respect to composition (the full dispersions are shown in the S.I. Figure 1).  
The magnitude and sign of the mode Gr\"uneisen parameters ($\gamma_i$ = -$\frac{V}{\omega_i}\frac{\delta\omega_i}{\delta V}$, where $V$ is volume and $\omega_i$ is the mode frequency), are represented by the thickness and color of the  dispersion curves.  Since the values are all scaled with a common factor, the curve thickness is representative of the relative values of the Gr\"uneisen parameter. 

We note that the predicted slope (i.e., velocity) and volume-dependence of the longitudinal acoustic branches are similar in CaMg$_2$Sb$_2$ and Mg$_3$Sb$_2$.  In contrast, the transverse phonon modes are much softer (lower velocity) and have a stronger volume dependence in Mg$_3$Sb$_2$.  The mode Gr\"uneisen parameters of the transverse phonons in Mg$_3$Sb$_2$ are particularly large at the Brillouin zone edge at the $A$-, $L$-, and $M$-point.  Those modes with the largest magnitude of $\gamma_i$ involve the shearing displacement of adjacent anionic slabs in the structure.  This is illustrated in Figure \ref{fig:displacement}, which shows snapshots of the displacement in the Mg$_3$Sb$_2$ structure for transverse acoustic phonons at the $A$-point (largest positive $\gamma_i$) and at the $L$-point (largest negative $\gamma_i$). 

By averaging the values of $\gamma_i$ over all modes, weighted by the heat capacity, we can estimate bulk Gr\"uneisen parameters of $\gamma$=1.83 and $\gamma$=1.44 in Mg$_3$Sb$_2$ and CaMg$_2$Sb$_2$, respectively.  However, as illustrated in Figure \ref{fig:gruneisen}, which shows the mode Gr\"uneisen parameters as a function of frequency, the highest mode Gr\"uneisen parameters in Mg$_3$Sb$_2$ are concentrated in the acoustic frequency range (highlighted in blue), which implies that they will have a large impact on thermal transport.  We note also, that the values of $\gamma_i$ in the optical frequency range from 120 to 150 cm$^-1$ are higher in Mg$_3$Sb$_2$ than in CaMg$_2$Sb$_2$.  These phonon modes involve almost exclusively the displacement of the cation (Mg(1) or Ca), further indication of unstable Mg(1)-Sb interlayer bonding.  Results for CaMg$_2$Bi$_2$, shown in the S.I. Figure 2, are similar to CaMg$_2$Sb$_2$.   

%comments from Toberer:
% Add: overall Grueneisen parameter is 1.8 vs 1.5 
% consider plotting grueneisen^2

%%%%%%%%%%%%%%%%%%%%%%%%%%%%%%%%%%%%%%%%%%%%%%%%%%%%%%%%%%%%%%%%%%%%%%%%%%%%%%%%%%%%%%%%%%%%%%%%%%%%%%%%%%%%%%%%%%%%%%%%%%%%%%%%%%%%%%%%%%%%%%%%%%%%%%%%%%%%%%%%%%%%%%%%%%%%%%%%%%%%%%%%%%%%%%%%%%%%%%

\subsection{Experimental elastic properties and thermal expansion}

The speed of sound and the elastic moduli are determined solely by the slope of the acoustic branches of the phonon dispersion at the $\Gamma$-point.  Thus, measuring the high-temperature elasticity offers a window into the behavior of the acoustic phonons near the $\Gamma$-point as a function of both temperature and unit cell volume, and can be used to evaluate the anharmonicity of these critical phonons. In the present study, resonant ultrasound spectroscopy (RUS) was used to obtain the elastic moduli, C11 and C44, of polycrystalline $A$Mg$_2$Pn$_2$ samples ($A$=Mg, Ca, Yb, and $Pn$=Sb, Bi) as a function of temperature, from which we obtain the shear and bulk moduli ($G$ and $K$, respectively) and the transverse and longitudinal speed of sound ($v_T$ and $v_L$, respectively). The details of the equations applied are listed in the S.I section 1. This method provides an accurate and non-destructive approach that has been used to study high-temperature elastic behavior of various classes of materials \cite{li2010high,astm1876standard}. Note that the elastic moduli and sound velocities obtained from polycrystalline samples represent an average over all crystallographic directions.    

The elastic moduli of solids tend to become softer with increasing bond length \cite{zeier2017new}.  Within compounds in the same structural pattern, if the unit cell volume increases, the elastic moduli are therefore expected to decrease.  As shown in Figure \ref{fig:elastic_moduli}, this trend is observed in $A$Mg$_2$$Pn$$_2$ compounds in both the experimental ($A$=Mg, Ca, Yb and $Pn$=Sb, Bi) and computational elastic moduli ($A$= Mg, Ca, Sr, Ba and $Pn$=P, As, Sb, Bi) obtained from the Materials Project \cite{ceder2010materials}.   The only significant outliers are the shear moduli of Mg$_3$Sb$_2$ and Mg$_3$Bi$_2$, which are much softer than compounds with similar unit cell volume. This suggests that the Mg(1)-Sb bonds connecting neighboring layers are quite weak compared with other $A$Mg$_2$$Pn$$_2$ compounds.   Note that we omitted the computed elastic moduli of rare-earth-containing compounds due to poor agreement with experiment. These values are included in the S.I.table 1, however.  Soft shear moduli are often seen in layered compounds with weak van der Waal bonds \cite{jenkins1972elastic,grimsditch1983shear,zabel2001phonons}
and in the extreme case, disappearing shear modes at high temperature have been associated with liquid-like behavior \cite{li2018liquid}.  
In layered Zintl phases, however, the ionic bonds between adjacent anionic layers are expected to be strong, particularly in comparison with van der Waals solids.   This is reflected by the relatively isotropic phonon dispersions of CaMg$_2$Sb$_2$ and CaMg$_2$Bi$_2$.  The soft shear moduli in Mg$_3$Sb$_2$ and Mg$_3$Bi$_2$, in contrast is anomalous.

\begin{figure}[!ht]
\centering
\includegraphics[width=0.48\textwidth]{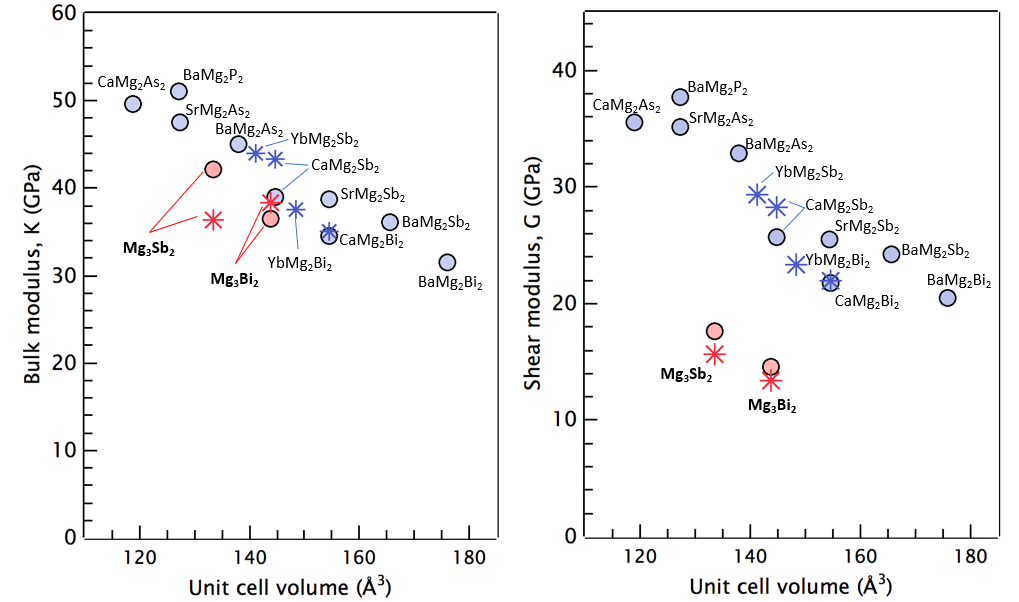}
\caption{Experimental (asterisk) and computed (circles) elastic moduli of $A$Mg$_2$Pn$_2$ compounds tend to decrease as a function of increasing unit cell volume and bond length.  The anomalously low shear moduli of Mg$_3$Sb$_2$ and Mg$_3$Bi$_2$ are significant, suggesting soft bonding unique to these two binary compounds \cite{de2015charting}.}
\label{fig:vibration_modes}
\end{figure}

\begin{figure*}[!ht]
    \centering
    \includegraphics[width=0.75\textwidth]{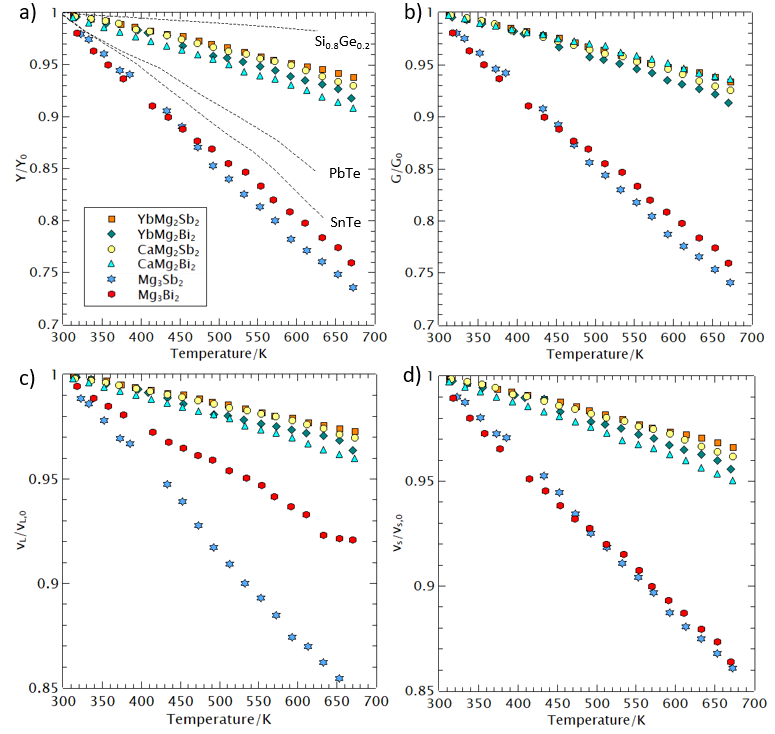}
    \caption{Temperature-dependent a) Young's modulus, b) shear modulus, and c) longitudinal and d) transverse speed of sound measured using resonant ultrasound spectroscopy.  Quantities were normalized to the room temperature value. Data for Si$_{0.8}$Ge$_{0.2}$, PbTe, and SnTe are from Ref. \cite{ren2008high,schmidt2013high,li2010high}.}
\label{fig:elastic_moduli}
\end{figure*}

Fig. \ref{fig:elastic_moduli} a) and b) shows the experimental temperature-dependence of the Young's and shear moduli of $A$Mg$_2Pn_2$ samples with $A$=Mg, Ca, Yb and $Pn$=Sb, Bi.  We have included the high-temperature Young's modulus of Si$_{0.8}$Ge$_{0.2}$ \cite{li2010high}, PbTe \cite{ren2008high} and SnTe \cite{schmidt2013high} for comparison.  In a purely harmonic model, the elastic moduli do not soften with increasing temperature. The degree to which a material deviates from this behavior can be used to quantify the degree of anharmonicity.   Si$_{0.8}$Ge$_{0.2}$ softens slowly with respect to temperature, consistent with its small Gr\"uneisen parameter ($\gamma$=1.06 in pure Si \cite{morelli2006high}), while PbTe and SnTe, which are known to be highly anharmonic, soften more rapidly ($\gamma$=2.1 in PbTe \cite{ren2008high}). Over the measured temperature range,we find that the elastic moduli of Mg$_3$Bi$_2$ and Mg$_3$Sb$_2$ soften by $\sim$ 25\%, in comparison to a $\sim$ 5\% decrease for CaMg$_2Pn_2$ and YbMg$_2Pn_2$ samples. The rapid rate of softening of the elastic moduli with increasing temperature ($\partial G/\partial T$ and $\partial Y/\partial T$) compared even with PbTe and SnTe, provides direct evidence of the high anharmonicity in the acoustic branches of these compounds.  

%The calculated rate of softening of the elastic moduli $\partial G/\partial T$ and $\partial Y/\partial T$ is predicted to be higher in the binary compounds, however, it is not as high as the experimental values.  Further, DFT predicts a more rapid decrease of the longitudinal speed of sound compared with that of the transverse speed of sound, while experimentally, we observe that both decrease at a similar rate. 

The thermal expansion coefficients ($\alpha$) of the compounds in this study were measured using high temperature X-ray diffraction (shown in the S.I.). 
%SI figure #??
For comparison, the computed thermal expansion as well as the temperature-dependent elastic moduli of Mg$_3$Sb$_2$, CaMg$_2$Sb$_2$, and CaMg$_2$Bi$_2$ are shown in the S.I Figure 3.  
Experimentally, we find that Mg$_3$Sb$_2$ and Mg$_3$Bi$_2$ have higher thermal expansion than other $A$Mg$_2Pn_2$ ($A$=Ca, Yb and $Pn$=Sb, Bi) compounds, as expected given their higher Gr\"uneisen parameters.  However, the contrast between the measured values of $\alpha$ in the binary versus the ternary variants is much less impressive than the difference in the temperature-dependence of their elastic moduli, $\partial G/\partial T$ and $\partial Y/\partial T$.  This likely reflects the stronger dependence of the $\partial G/\partial T$ and $\partial Y/\partial T$ on the acoustic branches, which, as shown above, are more anharmonic than the optical branches.  We thus regard the measurement of $\partial G/\partial T$ and $\partial Y/\partial T$ as an indicator of the ``acoustic Gr\"uneisen parameter''.

%The following relationship was derived by H. Ledbetter to relate the thermodynamic Gr\"uneisen parameter, $\gamma$, to the change in bulk modulus with respect to temperature, $\partial B/\partial T$ and the thermal expansion coefficient, $\alpha$:
%\begin{equation}
%\frac{1}{B}\frac{\partial B}{\partial T} = -(\gamma+1)\frac{1}{V}\frac{\partial V}{\partial T} = -(\gamma+1) \alpha
%\end{equation}
%where $B$ is the bulk modulus and $V$ is unit cell volume \cite{ledbetter1994relationship}.  This expression assumes that the material is isotropic. (Other assumptions??)
%While Mg$_3$Sb$_2$ and Mg$_3$Bi$_2$ have a higher thermal expansion coefficient than other $A$Mg$_2Pn_2$ ($A$=Ca, Yb and $Pn$=Sb, Bi) compounds, the $\sim$5:1 disparity in $\frac{1}{B}\frac{\partial B}{\partial T}$ is much greater than the roughly 1.5:1 difference in the thermal expansion coefficient. 
%Maybe we should put the slopes (v) and grueneisen in the SI....
%\textbf{Where to go with this? Something about this $\gamma$ being specific to the acoustic modes}.  We believe that the Gr\"uneisen parameter determined in this way reflects the mode $\gamma_i$ values of the acoustic phonon branches.

%extra text: Because unit cell volume and temperature increase simultaneously, the mode Grüneisen parameter, $\gamma_i$, cannot be extracted directly from the temperature dependence of the elastic moduli.

%%%%%%%%%%%%%%%%%%%%%%%%%%%%%%%%%%%%%%%%%%%%%%%%%%%%%%%%%%%%%%%%%%%%%%%%%%%%%%%%%%%%%%%%%%%%%%%%%%%

\subsection{Breaking Pauling's radius ratio rule}

\begin{figure}[!ht]
    \centering
    \includegraphics[width=0.4\textwidth]{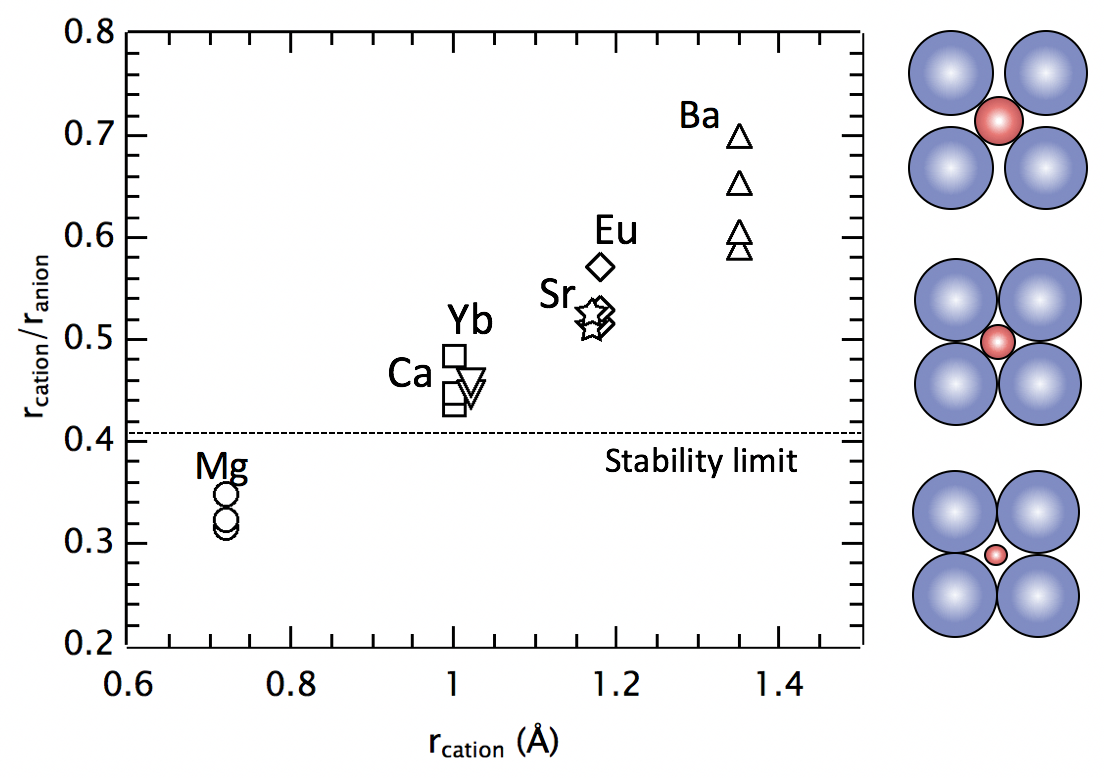}
    \caption{For octahedral coordination (CN=6), Pauling's radius ratio rules predict a minimum stability limit of $r_{cation}$:$r_{anion}$ = 0.414.  For Ca, Sr, Eu, and Yb, this rule is satisfied.  In contrast, the Mg cation is too small, which leads to a distorted octahedral environment and may be responsible for weak, anharmonic interlayer bonding. }
   \label{fig:radii_ratio}
\end{figure}

Among the $A$Mg$_2Pn_2$ compounds considered in this study, soft shear moduli and high anharmonicity appears to be unique to the binary compounds Mg$_3$Sb$_2$ and Mg$_3$Bi$_2$, despite their relatively low densities.  This begs the question of what is so special about the presence of Mg(1) on the octahedrally-coordinated cation site?  One immediately apparent factor is the size of Mg, which is smaller than any other cation that can occupy the octahedral site (e.g., Ca, Yb, Sr, Eu, and Ba). Several studies have investigated the effect of the cation size in $AM_2X_2$ compounds \cite{khatun2013quaternary,klufers1984alpha}, finding that the ThCr$_2$Si$_2$ structure type (in which the cation is 8-fold coordinated) is preferred over the CaAl$_2$Si$_2$ structure type in compounds with large cations such as Ba, K Rb.  Here, we consider the opposite extreme: cations that are too small.  In the sphere packing model proposed by Pauling for ionic solids, the smallest stable cation to anion radius for octahedral coordination is given by $r_{cation}$:$r_{anion}$=0.414 \cite{pauling1960nature}. Figure \ref{fig:radii_ratio} shows the estimated $r_{cation}$:$r_{anion}$ for $A$Mg$_2Pn_2$ compounds.  We employed ionic radii for the cations assuming a valence of 2+ and a coordination number of 6 from ref \cite{giacovazzo1992editor,shannon1969effective}.  The anionic radii were estimated empirically by taking the average $A$-$X$ distance in the CaAl$_2$Si$_2$ structure and subtracting the corresponding cation radii, yielding $r_P$ = 1.93 A, $r_{As}$ =2.07 A, $r_{Sb}$ = 2.23 A, and $r_{Bi}$ = 2.29 A.  Using this approach, compounds with $A$=Mg have $r_{cation}$:$r_{anion}$ below the stability limit, while compounds with larger cations are stable in a six-fold coordinated environment. 

It is arguable whether or not the use of the ionic $A$ radii is appropriate for this analysis, especially given the relatively small difference in electronegativity between Mg and Sb.  A comparison of the Born effective charges calculated for the Mg$_3$Sb$_2$ CaMg$_2$Sb$_2$ and CaMg$_2$Bi$_2$ (shown in the S.I. Table 3) would suggest that Mg(1)-Sb bonds are more covalent than Ca-Sb bonds. %Later we should add something about ELF calculations
If the Mg(1) effectively donates fewer than two valence electrons, its effective radius would also be larger, perhaps pushing the $r_{cation}$:$r_{anion}$ ratio above the predicted stability limit.  There are additional factors, however, supporting the view that Mg is unstable in the octahedral site, including the strong distortion of the octahedral environment around Mg and the presence of a phase transition at high temperatures.  The Mg$Pn_6$ octahedra in Mg$_3$Sb$_2$ and Mg$_3$Bi$_2$ exhibit bond angle variances of 26.53$^{\circ}$ and 36.01$^{\circ}$ respectively.  In comparison, the octahedral bond angle variance in CaMg$_2$Sb$_2$ and CaMg$_2$Bi$_2$ is only 1.77$^{\circ}$  and 5.04$^{\circ}$.  Further, Mg$_3$Sb$_2$ and Mg$_3$Bi$_2$ are the only $A$Mg$_2$Pn$_2$ compounds reported to undergo a structural phase transition at high temperature, which occurs at $\sim$900$^{\circ}C$ and $\sim$700$^{\circ}C$ respectively \cite{sevast2006binary}.  In the cubic, high-temperature structure (Sc$_2$O$_3$ structure type, space group $Ia3$), Mg is tetrahedrally coordinated and satisfies Pauling's radius ratio rule.  

The instability of the small Mg(1) cations in the octahedral $A$ site of the CaAl$_2$Si$_2$ structure is likely responsible for both the unusually soft shear moduli and the highly anharmonic transverse phonon modes in Mg$_3$Sb$_2$ and Mg$_3$Bi$_2$.  The weakening of ionic interlayer bonds would be expected to decrease the shear elastic moduli (Figure \ref{fig:elastic_moduli}, and thus reduce the speed of sound relative to stiff compounds.  As shown in Figure \ref{fig:kappa}b), this effect alone is insufficient to explain the factor of three reduction in $\kappa_L$ of Mg$_3$Sb$_2$ compared with CaMg$_2$Sb$_2$.  It is likely the undersized Mg(1) also has a destabilizing effect, leading to the observed high-temperature phase transition and also to the exceptionally high anharmonicity observed in low-frequency modes that are associated with shearing of adjacent layers, and in mid-frequency modes due to the motion of the Mg(1) cations (see Figure \ref{fig:dispersion} and Figure \ref{fig:displacement}).  ``Giant anharmonicity'' is well known to occur in the vicinity of lattice instabilities, as reported for PbTe, SnSe, GeTe,  \cite{hong2016electronic,li2015orbitally}, so it is not entirely surprising to see the same effect in this system as well.   Above the phase transition in Mg$_3$Sb$_2$ and Mg$_3$Bi$_2$, we would expect the elastic moduli, speed of sound, and lattice thermal conductivity to exhibit a sharp increase.  

In some ways, the undersized Mg in the octahedral site of the CaAl$_2$Si$_2$ structure is reminiscent of the filler atoms in clathrates and skutterudites which ``rattle'' in their oversized cages  \cite{nolas2001semiconductor,nolas1999skutterudites,uher2001skutterudites}. 
In such compounds, smaller cations are also desirable, as they can lead to stronger scattering and lower lattice thermal conductivity. However, cage compounds differ from the layered system in the present study in one important regard: the cages are formed from stiff covalent bonds that are not destabilized by undersized cations.  In the case of Mg$_3$Sb$_2$, the Sb atoms that form the "cage" around Mg have a closed shell configuration and cannot form bonds between adjacent layers. This destabilizing effect leads to highly anharmonic acoustic phonons.   In contrast, a fully filler rattler site leads to a flat optical mode coming down and crossing the acoustic modes \cite{toberer2011phonon}.
 
\section{Conclusion}

Inherently low lattice thermal conductivity is typically associated with dense materials or compounds with complexity at the atomic scale or on a micro-structural level.  The unusually low lattice thermal conductivity of Mg$_3$Sb$_2$, in contrast, shows that structural instability alone is sufficient, even in a very simple structure, to cause exceptionally high phonon scattering rates and low lattice thermal conductivity.  By combining ab initio phonon calculations and high-temperature elasticity measurements, we showed that both  Mg$_3$Sb$_2$ and Mg$_3$Bi$_2$ are highly anharmonic, in contrast to the Ca- and Yb-containing $A$Mg$_2Pn_2$ compounds investigated here.  Large mode Gr\"uneisen parameters, both negative and positive, were predicted in the acoustic branches of Mg$_3$Sb$_2$ and Mg$_3$Bi$_2$, which are expected to have a large contribution to thermal transport.  This was confirmed experimentally by the rapid decrease of the speed of sound and elastic moduli in Mg$_3$Sb$_2$ with increasing temperature, which is a direct consequence of the softening of the acoustic modes.  We attribute the unique behavior of Mg$_3$Sb$_2$ and Mg$_3$Bi$_2$ to the small radii of Mg, which is undersized for the octahedral coordination with Sb.  The poor fit of the Mg cation is suspected to lead to weak interlayer bonding, and thus to the observed soft shear modes, and ultimately, to the highly anharmonic behavior of the acoustic branches. These results suggest more broadly that soft shear modes resulting from undersized cations provide a potential path to low lattice thermal conductivity in ionic layered structures.

%extra text
%In particular, we see this as a potential strategy to be applied to layered or low dimensional Zintl phases in which the ionic bonds may be pushed to the point of destabilizing the structure.

\bibliography{Gruneisen}
\bibliographystyle{abbrv}
\end{document}